\documentclass[twocolumn,showpacs,preprintnumbers,amsmath,amssymb]{revtex4}
\usepackage{graphicx}
\usepackage{dcolumn}
\usepackage{bm}
\usepackage{color}
\usepackage[english]{babel}

\begin{document}

\title{
Nonanalyticity of the optimized effective potential with finite basis sets}
\author{Nikitas I. Gidopoulos$^{1,2}$, Nektarios N. Lathiotakis$^2$ \\
\em $^{1}$ISIS, STFC, Rutherford Appleton Laboratory, Didcot, OX11 0QX, United Kingdom\\
$^{2}$Theoretical and Physical Chemistry Institute, National Hellenic Research Foundation,
      Vass. Constantinou 48, GR-11635 Athens, Greece}
\date{\today}

\begin{abstract}
We show that the finite-basis optimized effective potential (OEP) equations exhibit previously unknown singular behavior.
Imposing continuity, we derive new well-behaved finite-basis-set OEP equations that determine OEP 
for any orbital and any large enough potential basis sets and which adopt an analytic solution via matrix-inversion.
\end{abstract}


\maketitle

\section{Introduction}

In the last couple of decades the optimized effective potential (OEP) theory 
\cite{sharp,talman,kummel2008,bl} appeared to offer a very promising route for 
improved accuracy in density functional theory (DFT) \cite{hk,ks,mel}. 
With OEP, not only is the exchange potential determined exactly but also
there is hope that correlation will eventually be approximated accurately, 
via an implicit density functional \cite{engel2006,rinke2005,pnas}. 
Furthermore, using OEP and imposing physical constraints on the 
potential, it is possible to improve the performance of traditional 
approximations, such as the local density approximation \cite{clda}.

Recently, it was discovered that finite basis implementations of OEP are marred 
by mathematical problems \cite{staroverov2006}. 
Several attempts have been made to overcome these issues \cite{hirata2001,gorling2008,hess,theo_vit,filatov1,filatov2,glushkov2009,thb} 
but with limited success so far. As a result and despite promise, interest in OEP has diminished.

The OEP is determined by a Fredholm integral equation of the first kind,
\begin{equation} \label{oepeqn0}
\int d{\bf r}' \chi_v ( {\bf r}, {\bf r}') v({\bf r}') = b_v ({\bf r}) \, .
\end{equation}
In this work, the OEP $v({\bf r})$ represents the sum of the Hartree 
and exchange-correlation 
potentials in the effective single-particle Hamiltonian, 
\begin{equation} \label{sp}
h_v({\bf r}) = -{ \nabla^2 \over 2} + v_{\rm en}({\bf r}) + v({\bf r}) \,,
\end{equation} 
where $v_{\rm en}({\bf r})$ is the electron-nuclear attractive potential. 
The single-particle eigenfunctions of $h_v$ and their energies build the density-density response function in (\ref{oepeqn0}), 
\begin{equation} \label{01}
\chi_v ( {\bf r} , {\bf r}') = 2 \sum_{i , a } { \phi_{v , i} ({\bf r}) \, 
\phi_{v , a} ({\bf r}) \, \phi_{v , a} ({\bf r}') \, \phi_{v , i} ({\bf r}') \over 
\epsilon _{v , i} - \epsilon _{v , a} } \, .  
\end{equation} 
Indexes $i$ and $a$ run respectively over occupied and unoccupied orbitals in the OEP Slater determinant.
$b_v ({\bf r})$ in (\ref{oepeqn0}) has units of density.
For concreteness, we omit correlation and focus on exchange OEP (x-OEP), where
\begin{equation} \label{02}
b_v({\bf r}) = 2 \sum_{i , a } { \langle a | {\cal J}_v  - {\cal K}_v | i \rangle 
\over \epsilon _i - \epsilon _a } \, \phi_{v , i} ({\bf r}) \, \phi_{v , a} ({\bf r}) \, ,
\end{equation}
where ${\cal J}_v({\bf r})$ is the direct Coulomb (or Hartree) local potential operator 
and ${\cal K}_v$ is the Coulomb exchange nonlocal operator.
We use the shorthand $\langle i | \cdot | a \rangle $ for the matrix element
$\langle \phi_{v, i} | \cdot | \phi_{v, a} \rangle$.

Hirata {\it et al.} \cite{hirata2001} proved that, for a complete orbital basis set, 
including continuum states, the products  
$\phi_{v , i}({\bf r}) \, \phi_{v , a} ({\bf r})$
form a complete set bar a constant. 
Hence, the response function (\ref{01}) is defined over the whole vector space of functions orthogonal to 
the constant function. Assuming that the response function is invertible in its space of definition, 
i.e. that the null-space of $\chi_v$ is only the constant function, 
we conclude that OEP is fully determined up to a constant.

On the other hand, with a finite orbital basis set 
the straightforward search for OEP may yield an infinity of solutions \cite{staroverov2006}:  
from Eqs.~(\ref{oepeqn0})-(\ref{02}) follows that the potential is undetermined in the null-space of 
the response function. 
The latter space contains only the constant function in the case of full $\chi_v$.
With a finite orbital basis set the infinite sum over virtual orbitals 
in Eq. (\ref{01}) is restricted to a sum over a subset of the virtual orbitals, 
in particular only over those obtained by the orbital basis functions.
The truncated response function $\chi_v^0$ obtained in this way has an infinite-dimensional null-space.
This results in indeterminacy of the potential along any of its components that lie in the null-space of $\chi_v^0$.  

Interestingly, several approximations of the finite-basis OEP approach \cite{kli,lhf,ceda,elp1,elp2}, 
which invariably employ the Uns\"old approximation \cite{ceda0}, 
determine the approximate OEP fully. 
The Uns\"old approximation amounts to a common energy denominator approximation 
for the static orbital Green's function, together with the closure or completeness relation. 
The reason is that with the closure relation 
the orbital basis set becomes effectively complete and the null-space of $\chi_v$ reduces 
once more to the constant function. 
Consequently it is no longer possible for an auxiliary basis function to have a component in the null-space of $\chi_v$. 

Returning to the OEP, the remedy appears readily, at least in principle. 
Consider a finite or even a complete auxiliary basis set 
$\{ \xi_n ({\bf r}) \}$ for the expansion of the potential and take 
the matrix elements of the truncated response function $\chi_v^0$ with the auxiliary basis functions:
\begin{equation} \label{akn}
A_{kn} \doteq - {1 \over 2} \, 
\iint d{\bf r} \, d{\bf r}' \, \xi_k ({\bf r} ) \, \chi_v^0 ({\bf r} , {\bf r}' ) \, \xi_n ({\bf r}' ) \, .
\end{equation} 
This matrix gives the projection of $\chi_v^0$ in the space spanned by the auxiliary basis functions. 
%
Its diagonalization 
yields eigenfunctions with singular and with 
nonsingular eigenvalues, which provide a convenient 
orthonormal basis in the auxiliary space. 
The null eigenfunctions of $A_{kn}$ are auxiliary functions that  
belong in the null-space of $\chi_v^0$ and cannot be used for the expansion of the potential, 
since the potential is undetermined along them.
The expansion of the potential along the remaining nonsingular eigenfunctions is well defined. 

\begin{figure} [h]
\vspace{0.4cm}
\includegraphics [width=0.95\columnwidth]{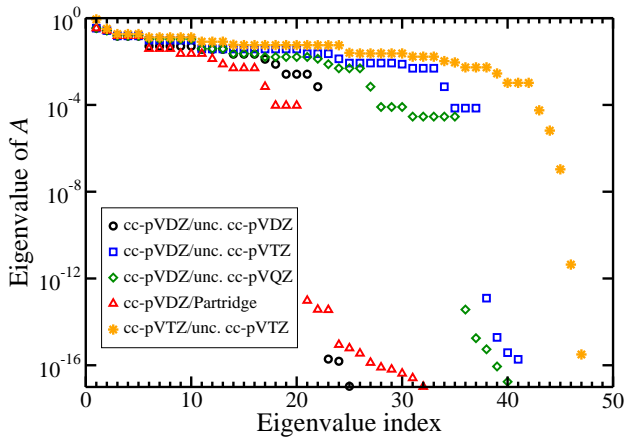} 
\caption{{The eigenvalues of matrix $A$ 
for the Ne atom. Circles, squares, diamonds, and triangles correspond to the cc-pVDZ orbital basis set and uncontracted cc-pVDZ, 
cc-pVTZ, cc-pVQZ and Partridge auxiliary basis sets respectively. 
Stars correspond to cc-pVTZ orbital and uncontracted cc-pVTZ auxiliary basis sets.}
} \label{f1}
\end{figure}

Unfortunately, such a singular value decomposition (SVD) and truncation of the null eigenvectors 
is in general ambiguous because the singular eigenvalues of the matrix $A_{kn}$ are not 
always separated unambiguously from the nonzero ones, as can be seen in Fig. \ref{f1} for the 
correlation-consistent polarized-valence triple-zeta
(cc-pVTZ) basis sets. 
Then, the inversion of $A_{kn}$ is not straightforward.

\begin{figure}
\vspace{0.4cm}
\includegraphics [width=0.95\columnwidth]{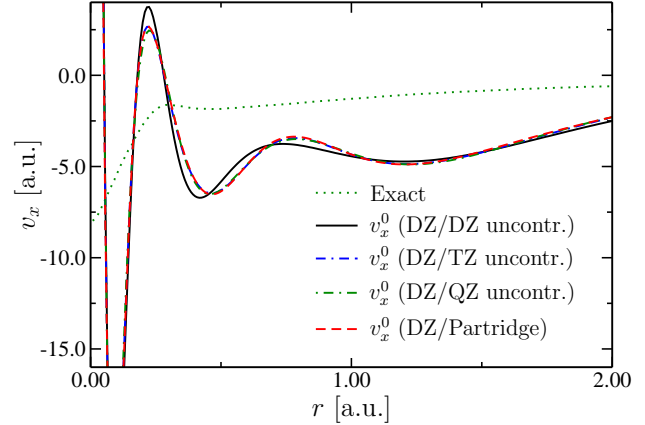} 
\caption{
Finite-basis exchange OEPs $v_{\rm x}^0$ for the Ne atom with the same cc-pVDZ orbital
     basis set and four different auxiliary basis sets:
uncontracted cc-pVDZ, uncontracted cc-pVTZ,
uncontracted cc-pVQZ, and uncontracted Partridge. The four potentials converge with an expanding auxiliary basis set 
to a potential that depends on the orbital basis only.  
The full numerical result from Ref. \cite{hess} is shown with a dotted line.}
\label{f1b}
\end{figure}

Nevertheless, sometimes, the singular eigenvalues of $A_{kn}$ can be unambiguously identified 
{\it a priori}, with a clear gap of many orders of magnitude separating them from the rest.
Such a case is shown in Fig.~\ref{f1} with the 
correlation-consistent, polarized-valence double-zeta
(cc-pVDZ) orbital basis set and several auxiliary basis sets. 
In this case, the resulting potential after truncation of the singular eigenvectors and inversion of $A_{kn}$ is 
expected to be unique. 
However, it is a mystery why in these cases the resulting potential, 
which is mathematically unique and well defined, looks unphysical with strong 
oscillations appearing near the nuclei; in the case of atoms, 
these oscillations make the potential look very different from the 
exact x-OEP \cite{hess}, as can be seen in Fig. \ref{f1b}. 

In Fig. \ref{f1b} we show the x-OEPs for the same orbital basis set cc-pVDZ and four 
auxiliary basis sets cc-pVDZ, cc-pVTZ, cc-pVQZ and Partridge (all uncontracted). 
As expected, since the truncation of the null eigenvectors of $A_{k n}$ is clear and straightforward, 
the four potentials in Fig. \ref{f1b} are close to each other and converge with an expanding auxiliary basis set. 
Thus, Fig. \ref{f1b} provides a numerical demonstration of our argument that when the SVD of matrix $A_{kn}$ is unambiguous, 
the resulting potentials from finite-basis OEP theory are mathematically well defined. 
Of course, it is surprising that the plots converge to a potential that looks unphysical and 
is significantly different from the full numerical x-OEP.

So far in the literature, the underlying reason for this anomaly has remained elusive 
and it is usually confused with the ill-posedness of the inversion of $A_{k n}$, 
the two problems appearing as one. 
For example, it is often written that similar wild oscillations of the potential near the nuclei appear 
when the orbital and auxiliary basis sets are not balanced. 
The lack of precise definition of balanced basis sets notwithstanding, 
we have observed in Figs.~\ref{f1} and \ref{f1b} that the anomaly is present after the truncation 
of auxiliary functions in the null-space of $\chi_v^0$, even 
in cases where the SVD of $A_{k n}$ is clear and unambiguous.

In this work, we distinguish between the cause for the anomalous behavior of the 
otherwise mathematically well-defined finite-basis OEP,
and the general ill-posedness of the choice of SVD cutoff for the truncation and inversion of $A_{k n}$.
We shall first focus on the former clear-cut problem which we shall resolve. 
Then, we shall discuss briefly the more common and complicated case, 
where the anomaly appears entangled with the ill-posedness of the inversion of the matrix $A_{k n}$ 
of the response function. 
We shall investigate this problem at length in a future publication. 
However, preliminary results presented in the last section 
indicate that our treatment of the nonanalyticity helps to sort out this problem as well. 

\section{analysis}

We shall base our analysis on the solution of the intermediate x-OEP equation, 
\begin{equation}
\chi_u \, v_{oep}=b_u \, , 
\end{equation}
with a fixed potential $u$ in $h_u$ [Eq.~(\ref{sp})],
i.e., $u$ is not updated to self-consistency; 
we also omit the explicit dependence on $u$. 
For simplicity, we shall analyze the consequence of a finite orbital basis set 
in the special case where the finite orbital set is composed of the occupied orbitals 
$\phi_i$ and of a subset of the (mostly lower lying) virtual orbitals $\phi_a$ of $h$.
The virtual orbitals of $h$ [Eq.~(\ref{sp})] that are outside the orbital basis 
form the complement basis; they will be denoted by 
$\tilde \phi_a$ (or simply $\tilde a$) and their energies will be denoted by $\tilde \epsilon _a$. 

In the first part of our analysis we consider that the potential is described on a grid 
with arbitrary precision, or equivalently that the potential is expanded in a complete auxiliary basis set. 

To study the effect of the finite orbital basis we split $\chi$ and $b$ in Eqs. 
(\ref{oepeqn0})-(\ref{02}): %
\begin{equation}
\chi ({\bf r} , {\bf r}' ) = \chi _{0} ({\bf r} , {\bf r}' ) + \tilde \chi ({\bf r} , {\bf r}' ) \, , 
\end{equation}
\begin{equation}
b({\bf r}) = b_0 ({\bf r}) + \tilde b ({\bf r}) \, , 
\end{equation}
where $\chi _{0}$ and $b_0$ are given by Eqs.~(\ref{01}) and (\ref{02}) 
but the sums over virtual states are restricted in the orbital finite basis.
The remainder functions (with a tilde) are defined by
\begin{eqnarray}
\tilde \chi ( {\bf r} , {\bf r}') &=& 2 \sum_{i , {\tilde a} } { \phi_{i} ({\bf r}) \, 
\tilde \phi_{a} ({\bf r}) \, \tilde \phi_{a} ({\bf r}') \, \phi_{i} ({\bf r}') \over 
\epsilon _{i} - \tilde \epsilon _{a} } \, , \\
\tilde b({\bf r}) &=& 2 \sum_{i , \tilde a } { \langle \tilde a | {\cal J}  - {\cal K} | i \rangle 
\over \epsilon _i - \tilde \epsilon _a } \, \phi_{i} ({\bf r}) \, \tilde \phi_{a} ({\bf r}) \, .
\end{eqnarray}
The summation over virtual orbitals is in the complement basis. 

Let us denote by $v^{\lambda}$ the potential which satisfies the equation, 
for $\lambda \ge 0$ (we drop the subscript $oep$),
\begin{equation} \label{oepeqn}
\left( \chi _{0} + \lambda \, \tilde \chi \right) 
v ^ \lambda = b_0 + \lambda \, \tilde b \, .
\end{equation}
The exact x-OEP is obtained for $\lambda =1$. 
In a finite-basis-set implementation the unknown $ \tilde \chi $ is always omitted, 
which amounts to setting $\lambda = 0$. 
Equation (\ref{oepeqn}) reduces to 
\begin{equation} \label{oepeqna}
\chi _{0} \, v ^ 0 = b_0 \, , 
\end{equation}
with 
\begin{equation} \label{oepeqn00}
v^{ 0} = \chi_0^{-1} \, b_0 
\end{equation}
in place of $v^{ 1}$, representing the finite-basis x-OEP solution. 

However, we point out that for finite $\lambda > 0$, the response function $( \chi _{0} + \lambda \, \tilde \chi )$ 
in (\ref{oepeqn}) is invertible and $v^\lambda$ is fully determined up to a constant.
On the other hand, for $\lambda = 0$ the response function reduces to $\chi_0$, which 
has an infinite-dimensional null-space where $v^{0}$ is undetermined. 
Hence, the complete omission of $\tilde \chi $ is a singular operation and the following theorem holds:

{\em Theorem 1}. The solution $v^{\lambda}$ of the OEP equation (\ref{oepeqn}) 
is not a continuous function of $\lambda$ at $\lambda =0$,
\begin{equation} \label{sing}
v^{\lambda \rightarrow 0} \ne v^{0} \, ,
\end{equation}
where $v^{\lambda \rightarrow 0}$ stands for $\lim_{\lambda \rightarrow 0} v^{\lambda} $.


To expand on the proof of the theorem, we consider the eigenfunctions $c^\alpha ({\bf r})$ of $\chi_0$ with nonzero 
eigenvalues and an orthonormal basis $c^\nu ({\bf r})$ in the null-space of $\chi_0$:
(i.e. $g^\nu = 0$)
\begin{equation}
\int d{\bf r}' \, \chi_0 ({\bf r},{\bf r}') \, c^\alpha ({\bf r}')  = - 2 \, g^\alpha \, c^\alpha ({\bf r}) \label{eig} 
\end{equation}
The index $\alpha$ enumerates the nonsingular eigenfunctions 
and $\nu , \mu$ run over the null eigenfunctions.
The functions $c^\nu ({\bf r})$ cannot be chosen to be eigenfunctions of $\tilde \chi$ since the latter 
has a projection in the nonsingular space of $\chi_0$.    

We shall use the complete set of eigenfunctions of $\chi_0$ as a special basis in which we expand the potential.  

From Eqs. (\ref{oepeqna}) and (\ref{oepeqn00}), the potential 
$v^0 ({\bf r})$ is expressed in terms of the nonsingular eignfunctions $c^\alpha ({\bf r})$: 
\begin{equation} \label{eqv0}
v^0 ({\bf r}) = \sum_\alpha { b_\alpha \, c^\alpha ({\bf r}) \over g^\alpha} \, .
\end{equation}
where $b_\alpha$ is the overlap:
\begin{equation} \label{17}
b_\alpha \doteq - {1 \over 2} \, \int d{\bf r} \, c^\alpha ({\bf r}) \, b_0 ({\bf r})  \, . 
\end{equation}
We also define the overlap:
\begin{equation}
{\tilde b}_\nu \doteq - {1 \over 2} \, \int d{\bf r} \, c^\nu ({\bf r}) \, \tilde b ({\bf r})  \, .
\end{equation}

We expand in Taylor series $v^\lambda$, around the value $v^\kappa$, with $\lambda > \kappa > 0$: 
\begin{equation}
v^\lambda = v^\kappa + (\lambda - \kappa) \, v'_\kappa + {(\lambda  - \kappa ) ^2 \over 2} \, v''_\kappa + \ldots
\end{equation}
We take the limit $\kappa \rightarrow 0$. 
To allow for the possibility $v^{\kappa \rightarrow 0} \ne v^0$, 
we write:
\begin{equation} \label{1}
v^\lambda =  v^{\lambda \rightarrow 0} + \lambda \, v ' + {\lambda^2 \over 2} \, v'' + \ldots 
\end{equation}
To determine $v^{\lambda \rightarrow 0}$, 
we substitute (\ref{1}) in (\ref{oepeqn}), use (\ref{oepeqna}), and keep up to first order in $\lambda$: 
\begin{equation} \label{8}
\chi_0 \, v^{\lambda \rightarrow 0} + \lambda \, ( \chi_0 \, v' + \tilde \chi \, v^{\lambda \rightarrow 0} ) = \chi_0 \, v^0 + \lambda \, \tilde b\,.
\end{equation}
From the zero-order equation, $\chi_0 \, v^{\lambda \rightarrow 0}  = \chi_0 \, v^0$, we obtain that $ \bar v \doteq v^{\lambda \rightarrow 0} - v^0 $ 
is a null vector of $\chi_0$ and can be expanded in the null eigenfunctions of $\chi_0$, 
\begin{equation}
{\bar v} ({\bf r}) = \sum_\mu {\bar v}_\mu \, c^\mu  ({\bf r}) \,.
\end{equation}
The potential $\bar v$ is a measure of the discontinuity. 

The linear term in Eq. (\ref{8}) gives: 
\begin{equation}
\chi_0 \, v' + \tilde \chi \, v^0 + \tilde \chi \, \bar v = \tilde b \,. \label{5}
\end{equation}
Multiplying by $c^\nu ({\bf r})$, integrating over $\bf r$, 
and using 
\begin{equation}
\int d{\bf r} \,  c^\nu ({\bf r} ) \, \chi_0 ({\bf r} , {\bf r} ') = 0 \, , 
\end{equation}
we obtain,
\begin{equation} 
\sum_\mu {\tilde \chi}_{\nu \mu} \, {\bar v}_\mu = 
{\tilde b}_\nu - \sum_{ \alpha }  {\tilde \chi}_{\nu \alpha} \, {b_\alpha \over g^\alpha}  \, , 
\end{equation}
where, 
\begin{eqnarray} 
{\tilde \chi}_{\nu \mu} & \doteq & - {1 \over 2} \, 
\iint d{\bf r} \, d{\bf r}' \, c^\nu ({\bf r} ) \, \tilde \chi ({\bf r} , {\bf r}' ) \, c^\mu ({\bf r}' ) \, , \\
{\tilde \chi}_{\nu \alpha } & \doteq &- {1 \over 2} \, 
\iint d{\bf r} \, d{\bf r}' \, c^\nu ({\bf r} ) \, \tilde \chi ({\bf r} , {\bf r}' ) \, c^\alpha ({\bf r}' ) \, .
\end{eqnarray}
The null-space of $\chi_0$ (where $c^\nu$ and $c^\mu$ lie) is a proper subset of the nonsingular space of 
${\tilde \chi}$ and hence the matrix ${\tilde \chi}_{\nu \mu} $ is well-defined
and invertible. Also, the nonsingular space of ${\tilde \chi}$ overlaps with the nonsingular space
of $\chi_0$ (where $c^\alpha$ lie) and the matrix ${\tilde \chi}_{\nu \alpha}$ does not vanish identically. 
We finally have:  
\begin{equation} \label{5a}
{\bar v}_\mu = 
\sum_\nu {\tilde \chi}^{-1}_{\mu \nu} \, {\tilde b}_\nu - \sum_{ \alpha , \nu} {\tilde \chi}^{-1}_{\mu \nu} \, 
{\tilde \chi}_{\nu \alpha} \, \, {b_\alpha \over g^\alpha} \, . 
\end{equation}
In general, the right-hand side and hence $\bar v$ are not expected to vanish. 
This completes the proof that $v^{\lambda}$ is discontinuous at $\lambda = 0$. QED 

Equation (\ref{5a}) determines $\bar v$ in the null-space of $\chi_0$. 
In order to calculate $\bar v$ from Eq. (\ref{5a}) we need 
some knowledge of ${\tilde \chi}$. 
In the next section we shall use the Uns\"old approximation to approximate 
${\tilde \chi}$.

To conclude this section, it is important to note that the nonanalyticity of 
OEP at $\lambda = 0$ is the result of the truncation of the orbital basis set. 
So far we have taken that the potential is either represented on a grid with arbitrary precision, 
or that it is expanded in terms of the complete set of eigenfunctions of $\chi_0$.
Therefore, the finiteness of the auxiliary basis set (in practice) has not played a role.
In the following, we shall expand the potential in an arbitrary but complete auxiliary basis set to
complete the derivation of the new finite-basis-set OEP equations in which the potential 
is not discontinuous at $\lambda = 0$. 

\subsection*{Arbitrary complete auxiliary basis set.} 

So far we have expressed the potential in the basis of the eigenfunctions $\{ c^\alpha ({\bf r}) \}$ of the truncated 
density-density response function $\chi_0 ({\bf r}, {\bf r}')$, complemented for completeness 
by an orthonormal basis $\{ c^\nu ({\bf r}) \}$ in the null-space of $\chi_0 ({\bf r}, {\bf r}')$.  
Although $\{ c^p ({\bf r}) \}$, ($p = \alpha , \, \nu$), has been a natural basis to base our analysis on, 
it is not a practical basis for calculations and in this section we shall make a change of 
basis in the auxiliary space and expand the potential in an arbitrary but complete auxiliary basis set $\{ \xi_n ({\bf r}) \}$. 
Of course, in practice, the auxiliary basis set is never complete; however, the finiteness of the auxiliary basis set 
is not expected to introduce any further nonanalyticity in the potential. 
In addition, the new finite-basis OEP equations, derived in this section, 
give meaningful results for a finite auxiliary basis
as long as the latter is large enough to overlap with the null-space of $\chi_0$.

In the following, we use a complete auxiliary basis set $\{ \xi_n \}$ to expand 
the eigenfunctions of the response function, 
\begin{equation} \label{29}
c^p ({\bf r})= \sum_{n} c^p _{n} \, \xi_{n}  ({\bf r}) \, , \ \ p = \alpha, \nu \, .
\end{equation}
Equation (\ref{29}) defines the transformation between the auxiliary bases: $\{ c^p ({\bf r}) \} \rightarrow \{ \xi_n ({\bf r}) \}$.
  
Substituting Eq. (\ref{29}) in the eigenvalue equation (\ref{eig}) and using (\ref{akn}), we obtain that the 
coefficients satisfy the generalized eigenvalue equations
\begin{eqnarray} 
\label{eigsa}
\sum_n A_{k n} \, c^\alpha_n & = & g^\alpha \, \sum_n \langle \xi_k | \xi_n \rangle \, c^\alpha_n \, , \\ 
\sum_n A_{k n} \, c^\nu_n & = & 0 \, ,  
\end{eqnarray}
where
$A_{k n}$ in Eq. (\ref{akn}) is given explicitly by 
\begin{equation}
A_{k n} = \sum_{i , a } { \langle i | \xi_{k} | a \rangle \langle a | \xi_{n} | i \rangle 
\over \epsilon _a - \epsilon _i }  \label{me1} \, . 
\end{equation}
To distinguish from the previous basis of the eigenfunctions of $\chi_0$, 
we use capital letters for the matrix elements of the response function 
with the auxiliary basis $\{ \xi_n ({\bf r}) \}$ and for the
overlap of the auxiliary basis functions with $b_0 ({\bf r})$ and $\tilde b ({\bf r})$, 
\begin{eqnarray}
\tilde A_{k n} & \doteq & - {1 \over 2} \, 
\iint d{\bf r} \, d{\bf r}' \, \xi_k ({\bf r} ) \, \tilde \chi ({\bf r} , {\bf r}' ) \, \xi_n ({\bf r}' ) \, , \\ 
B_k &\doteq &- {1 \over 2} \int \!\! d {\bf r} \, \xi_k ({\bf r}) \, b_0 ({\bf r}) \, , \label{34} \\
\tilde B_k & \doteq & - {1 \over 2} \int \!\! d {\bf r} \, \xi_k ({\bf r}) \, \tilde b ({\bf r}) \, .
\end{eqnarray}
From (\ref{02}) $B_k$ becomes explicitly
\begin{equation}
B_k = \sum_{i , a } { \langle i | \xi_{k} | a \rangle 
\langle a | {\cal J}  - {\cal K} | i \rangle \over \epsilon _a - \epsilon _i }  \label{me3} \, . 
\end{equation}

Obviously, we cannot have $\tilde \chi$ and $\tilde b$ exactly.
We approximate \cite{ceda0} the energy differences in the denominators of 
$\tilde \chi $ and $\tilde b$ by a constant, 
$\Delta \simeq \tilde \epsilon _{u,a} - \epsilon _{u,i}$ and then use the closure relation. 
The matrix elements become: 
\begin{equation}
\tilde A_{k n}  =  \sum_{i} \langle i | \xi_{k} \xi_{n} | i \rangle 
-\sum_{i,j} \langle i | \xi_{k} | j \rangle \langle j | \xi_{n} | i \rangle  \label{me2}
\end{equation}
\begin{equation} \label{me4}
\tilde B_{k }  = \sum_{i} \langle i | \xi_{k} ( {\cal J}  - {\cal K} ) | i \rangle  
-\sum_{i,j} \langle i | \xi_{k} | j \rangle \langle j | {\cal J}  - {\cal K} | i \rangle 
\end{equation}
In Eqs. (\ref{me2}) and (\ref{me4}) we have ignored a term in ${\tilde A}_{k n}$ 
and a term in $\tilde B_k$, whose contribution in the OEP equation below, 
in the limit $\lambda \rightarrow 0$, vanishes smoothly.

Next, we expand the potential in the new basis,  
\begin{equation}
v^\lambda  ({\bf r})= \sum_{n} v_{n}^\lambda  \,  \xi_{n}  ({\bf r}) \, .
\end{equation}
In matrix form the OEP equation (\ref{oepeqn}) becomes
\begin{equation} \label{nnoep}
\sum_n ( A_{k n} + \lambda \, \tilde A_{k n} ) \,  v_{n}^\lambda =  B_k + \lambda \, \tilde B_k  \, , 
\end{equation}

With definitions (\ref{me2},\ref{me4}), $\lambda$ in Eq.~(\ref{nnoep}) stands for $\lambda / \Delta $. 
In the limits $\lambda \rightarrow 0$ and $\lambda \rightarrow \infty$ 
our results are independent of $\Delta$.   

For fixed $\lambda$ the solution of Eq.~(\ref{nnoep}) is 
\begin{equation} \label{nnoep1}
v_{n}^\lambda = \sum_k ( A + \lambda \, \tilde A )_{n k}^{-1} \, ( B_k + \lambda \, \tilde B_k ) 
\end{equation}
For $\lambda \rightarrow \infty$, we observe that $v^\infty$ satisfies the effective
local potential (ELP) equations \cite{elp1,elp2}:
\begin{equation} \label{kli}
v_{n}^\infty = \sum_k \tilde A _{n k}^{-1} \, \tilde B_k \, .
\end{equation}
In the literature, the ELP potential $v^\infty$ is also known as common-energy-denominator approximation (CEDA)\cite{ceda}, or local Hartree-Fock (LHF)\cite{lhf}.

The same analysis, Eqs.~(\ref{1})-(\ref{5a}), which led us to conclude that the potential $v^\lambda$ in Eq.~(\ref{oepeqn}) 
is discontinuous applies to Eqs.~(\ref{nnoep}) and (\ref{nnoep1}). 
For small $\lambda$, the potential $v^\lambda$ tends to the limit 
\begin{equation} \label{23m}
v^{\lambda \rightarrow 0} = v^{0} + \bar v . 
\end{equation}
The two components $v^0$ and $\bar v$ are given by Eqs.~(\ref{eqv0}) and (\ref{5a}).
We expand $v^0$ and $\bar v$ in the auxiliary basis set,
\begin{equation} \label{43}
v^0 ({\bf r}) = \sum_n v^0_n \, \xi_n ({\bf r}) \, , \ \ {\bar v} ({\bf r}) = \sum_n {\bar v}_n \, \xi_n ({\bf r})
\end{equation}
The change of the auxiliary basis $ \{ c^p ({\bf r}) \} \rightarrow \{ \xi_n ({\bf r}) \}$ is given by Eq. (\ref{29}).
It is straightforward to obtain
\begin{eqnarray} 
v^0_n & = & \sum_\alpha c^\alpha_n \, v^0_\alpha\,, \\
{\bar v}_n & = & \sum_\mu c^\mu_n \, {\bar v}_\mu\,, \\
b_\alpha & = & \sum_k c^\alpha_k \, B_k\,, \\
{\tilde b}_\nu & = & \sum_k c^\nu_k \, {\tilde B}_k  \,.
\end{eqnarray}
Finally, Eqs.~(\ref{eqv0}) and  (\ref{5a}) become
\begin{equation} \label{230}
v^{0}_n =
\sum_{\alpha  }  {c^\alpha_n  \over g^\alpha} \sum_k c^\alpha_k \, B_k \, ,  
\end{equation}
\begin{equation}\label{23} 
\bar v_n  = \sum_{\mu ,\nu , k} c^\mu _n \, {\tilde \chi} ^{-1} _{\mu \nu } \,
 c^{\nu }_k \, \tilde B_k 
- \sum_{\alpha  , \mu , \nu }  c^\mu _n \, {\tilde \chi} ^{-1} _{\mu \nu } \, \tilde \chi_{ \nu \alpha} 
{ \sum_k c^\alpha_k \, B_k  \over g^\alpha}  \, .
\end{equation}
In contrast to Eq. (\ref{5a}), the matrices 
${\tilde \chi}_{\mu \nu}$ and ${\tilde \chi}_{\nu \alpha}$ above are not given exactly but within the Uns\"old approximation: 
\begin{equation} \label{}
{\tilde \chi } _{\mu \nu } 
 = \sum_{k , n} c^{\mu }_k \, \tilde A _{k n} \, c^{\nu  }_n \, , \ \ \ 
{\tilde \chi} _{\nu \alpha  } = \sum_{k , n} c^{\nu }_k \, \tilde A _{k n} \, c^{\alpha  }_n 
\end{equation}
Therefore, Eqs.~(\ref{230}) and (\ref{23}) determine $v^0$ exactly but $\bar v$ only approximately.

The finite-basis x-OEP equations~(\ref{23m}), (\ref{43}), (\ref{230}), and (\ref{23}) are the main result of this work. 
In these equations, the discontinuity of the potential $v^\lambda $ as a function of $\lambda$ at $\lambda = 0$ is evident.
Until now, the potential $v^{0}$ played the role of finite-basis x-OEP. It is obtained 
by truncating the singular eigenvectors of $A$ and subsequently inverting $A$ in the space of nonsingular
eigenvectors. 
The component $\bar v$ of $v^{\lambda \rightarrow 0}$ determines x-OEP in parts of the 
auxiliary space where $v^{0}$ is undetermined.
For example, the first term on the right-hand side of Eq. (\ref{23}) is the projection of $v_{}^{\infty}$ 
in the null-space of $A_{k n}$.
Also if the null-space of $A_{k n}$ artificially included all the eigenvectors of $A_{k n}$, then
from Eqs.~(\ref{kli}), (\ref{230}), and (\ref{23}), 
we would have $v^{0} = 0$ and $v^{\lambda \rightarrow 0} = \bar v = v_{}^{\infty}$.

The total energy as a function of $\lambda$ is continuous, i.e., the total energy of 
$v^{\lambda \rightarrow 0}$ is the same as that of $v^{0}$. 

\section{Numerical Examples}

\begin{figure} 
\vspace{0.4cm}
\includegraphics [width=0.95\columnwidth]{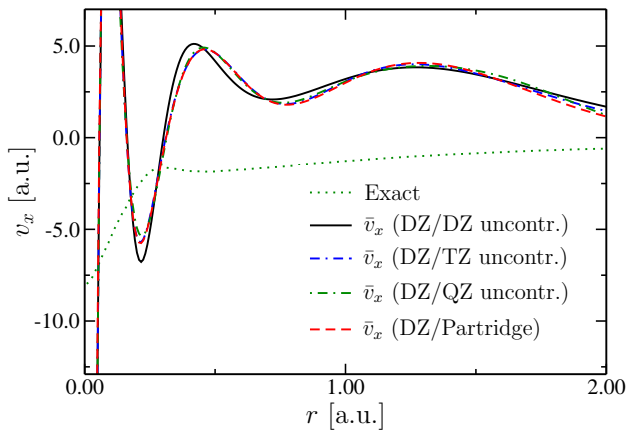}
\caption{
The exchange part, $\bar{v}_{\rm x} $ of the correction potentials $\bar{v}$, from Eqs.~(\ref{23m},~\ref{23}), corresponding to the 
potentials $v_{\rm x}^{0}$ shown in Fig.~\ref{f1b} for the Ne atom.
\label{fig:vbar}
}
\end{figure}

The eigenvalue spectrum of $A_{k n}$ for the Ne atom, 
using different basis set combinations for orbital and auxiliary basis 
sets, is shown in Fig.~\ref{f1}. In our calculations, we use Hartree-Fock (HF) orbitals in place 
of $\phi_{u,i}$, $\epsilon _{u , i}$, $\phi_{u,a}$ and $\epsilon _{u , a}$.
A gap separates the zero from the nonzero eigenvalues for
the cc-pVDZ orbital basis set and four auxiliary basis sets. 
In Fig.~\ref{f1b}, we show that the four different potentials $v^0$ converge with expanding auxiliary basis 
to a potential that is characteristic of the orbital basis only.
In Fig.~\ref{fig:vbar}, we perform the same test for the newly defined complementary potential $\bar v$. 
Again, we observe that the four different potentials $\bar v$ converge and the converged potential depends 
only on the orbital basis.

\begin{figure} [h]
\vspace{0.6cm}
\includegraphics [width=0.95\columnwidth]{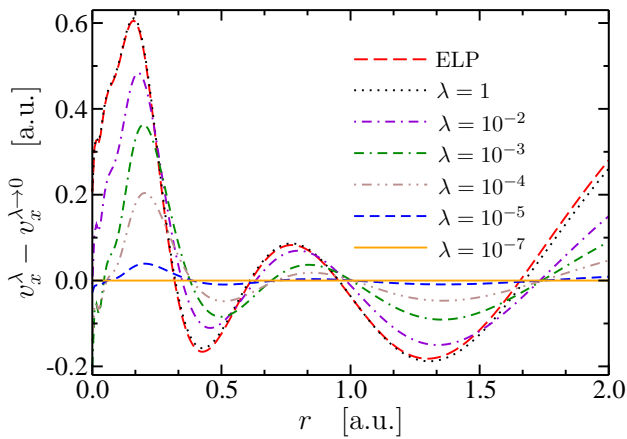} 
\caption{
Differences of the exchange potentials $v_{\rm x}^\lambda - v_{\rm x}^{\lambda \rightarrow 0}$ 
for various values of $\lambda$ for the Ne atom with cc-pVDZ and uncontracted cc-pVDZ for the 
orbital and auxiliary basis set respectively.
The potentials $v_{\rm x}^{\lambda}$ are calculated using Eq. (\ref{nnoep1}), and 
$v_{\rm x}^{\lambda \rightarrow 0}$ is calculated using Eqs.~(\ref{23m}), (\ref{43}, (\ref{230}, and (\ref{23}). 
The red long-dashed line is the difference $v_{\rm x}^{\infty} - v_{\rm x}^{\lambda \rightarrow 0}$, 
where $v_{\rm x}^{\infty} $ is calculated with Eq.~(\ref{kli}).
} 
\label{f10}
\end{figure}

In Fig.~\ref{f10}, (Ne atom, cc-pVDZ and uncontracted cc-pVDZ orbital and auxiliary basis sets), 
we investigate numerically the interpolation of the potential $v^{\lambda}$
between the limiting potentials $v^{\lambda \rightarrow 0}$ and $v^\infty$ 
by plotting the potential difference $v^{\lambda} - v^{\lambda \rightarrow 0}$. 
The smooth convergence of $v^{\lambda}$ toward the two limits for small and for large $\lambda$ can be seen clearly. 
On the scale of the plot, the potential $v^{\lambda}$ is on top of $v^{\infty}$ already at $\lambda=10$.  
For $\lambda=10^{-8}$ the potential $v^{\lambda}$ is indistinguishable from $v^{\lambda \rightarrow 0}$ and their difference vanishes.

The limiting potential $v^{\lambda \rightarrow 0}$ appears in 
Fig.~\ref{f2}(a) (cc-pVDZ and uncontracted cc-pVDZ for orbital and auxiliary basis sets), 
together with the potentials $v^{0.0001}$ and
$v^\infty$, 
and for reference the full numerical grid result from Ref.~\onlinecite{hess}.
The potential $v^{\lambda \rightarrow 0}$ lies almost on top of $v^{0.0001}$ and 
is quite close to the exact potential. 
Thus, by adding $\bar{v}$ to $v^{0}$,
the anomalous oscillatory behaviour of $v^{0}$ was corrected. 
This effect is further demonstrated in the inset of Fig.~\ref{f2}(a) where 
we show the two constituent potentials $v^{0}$ and $\bar v$ and their sum $v^{\lambda \rightarrow 0}$.
For this particular basis set, the total energy of x-OEP, for $\lambda = 0$ and $\lambda \rightarrow 0$, 
is identical to HF total energy. 

For each pair of orbital and auxiliary basis sets the plots of the potentials 
$v^0$ and $v^{\lambda \rightarrow 0}$ are shifted  by the same constant, which  
equals the difference between the highest occupied eigenvalues in the HF and the $v^{\lambda \rightarrow 0}$ 
OEP calculations. No shift is applied to the correction potentials $\bar v$. 

\subsection*{Ill-posed inversion of $A_{k n}$}

In general, the problem of how to separate the singular ($g^{\nu}$) from the 
nonsingular ($g^\alpha$) eigenvalues of $A_{k n}$ is considered not straightforward and ill-posed \cite{savin1,thb}. 
We focus on this problem in a forthcoming publication \cite{nn_tikh}, 
where we follow Ref. \cite{thb} and employ Tikhonov's regularization theory \cite{tikhonov}, 
after deriving an appropriate regularizing function for x-OEP, 
from the energy difference $T_u [v]$ in Ref. \onlinecite{pnas}.  
In the rest of the section we present a simpler and preliminary analysis.

An example where the logarithms of the eigenvalues of $A_{k n}$ spread almost uniformly 
between singular and nonsingular values is shown in Fig.~\ref{f1}
for the Ne atom for orbital and auxiliary basis sets cc-pVTZ and uncontracted cc-pVTZ, respectively.
In Fig.~\ref{f1}, we observe that the logarithms of the eigenvalues fall broadly in two groups with different slopes for 
increasing eigenvalue index $p$. 
In the first group, with $p \le 42$, the slope is not steep and the logarithms of the eigenvalues 
fall off rather slowly. In the second group, with $p \ge 43$ the logarithms of the eigenvalues fall off faster.

\begin{figure} [t]
\centering 
\vspace{0.5cm}
{
\includegraphics [width=0.95\columnwidth]{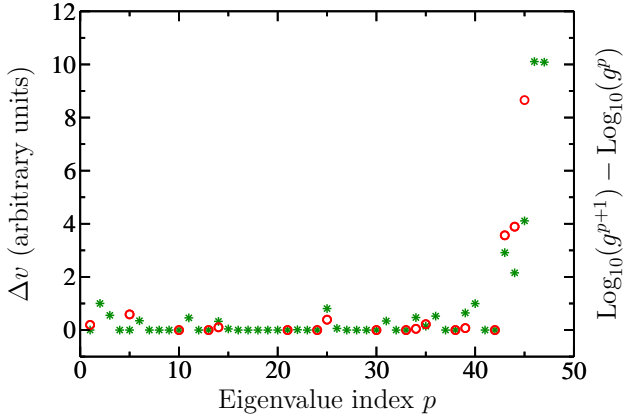} 
}
\caption{
The step change of potential $\Delta v$ (circles) and the quantity 
$\log_{10}{(g^{p + 1})} - \log_{10}{(g^{p})}$ (stars) vs the index $p$ of the 
eigenvalues (in order of decreasing magnitude) of the matrix $A_{k n}$ for Ne with cc-pVTZ and uncontracted cc-pVTZ orbital 
and auxiliary basis sets.
\label{fig:dv}} 
\end{figure}

In Fig.~\ref{fig:dv}, we plot the difference $\log_{10}{(g^{p + 1})} - \log_{10}{(g^{p})}$ of consecutive eigenvalues versus the eigenvalue index $p$.
We see there is a sharp increase in this difference as the eigenvalue index crosses the value from one group to the other.
The lowest eigenvalue in the first group is $g^{\alpha} = 1.04 \times 10^{-3}$.

\begin{figure*}[t] 
\begin{tabular}{cc}
\includegraphics [width=0.95\columnwidth]{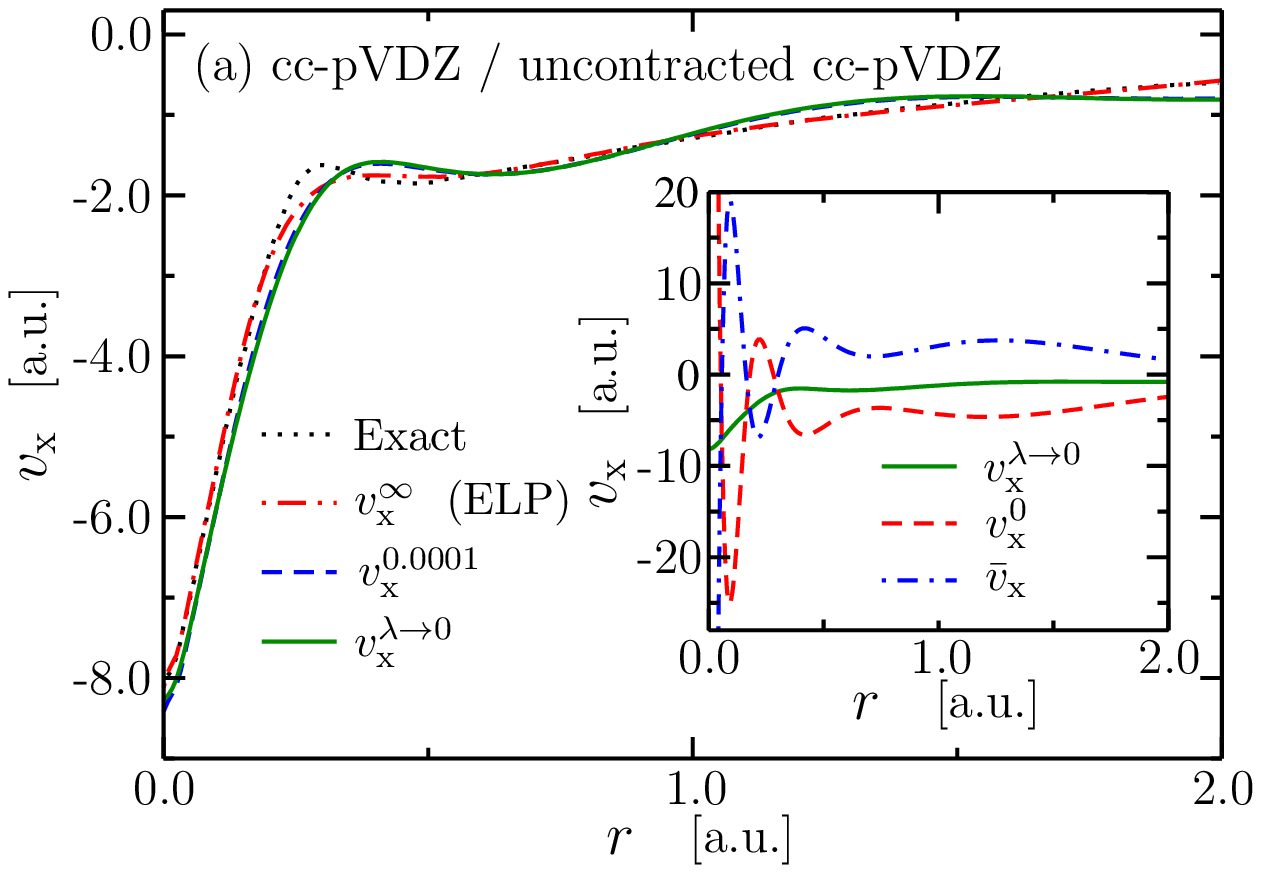} \ & \ 
\includegraphics [width=0.95\columnwidth]{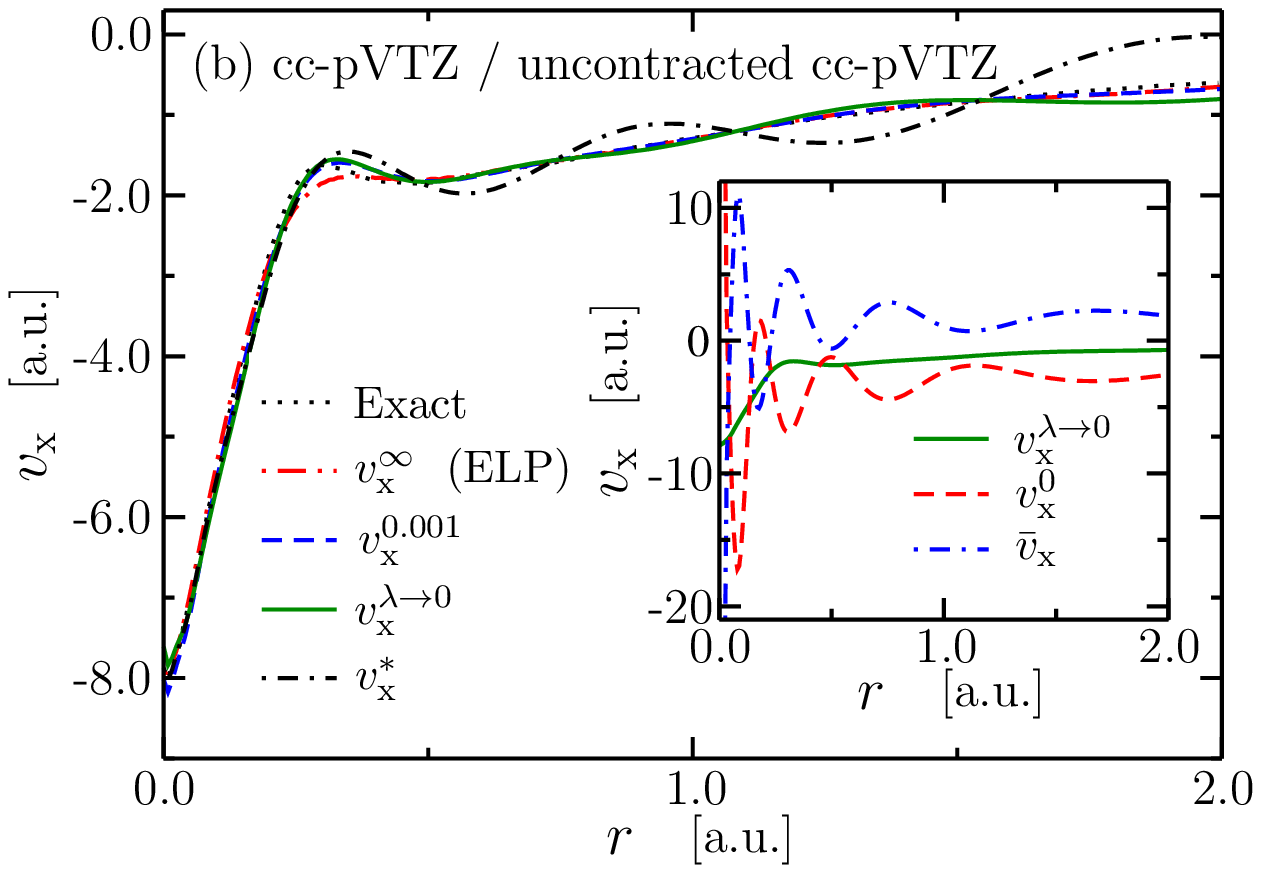} \\
\end{tabular}
\caption{ \label{f2}
Exchange potentials, for $\lambda \rightarrow {\infty}$, $\lambda \rightarrow 0$,
of the Ne atom using two different combinations for the orbital and auxiliary basis sets: (a) cc-pVDZ and cc-pVDZ 
uncontracted and (b) cc-pVTZ and uncontracted cc-pVTZ. The full numerical result from 
Ref.~\onlinecite{hess} on a grid is shown as exact. 
In the insets, the two strongly oscillating components of $v_{\rm x}^{\lambda \rightarrow 0}$ 
of Eq.~(\ref{5a}) are shown. Potentials for $\lambda =0.0001$ and $\lambda =0.001$ are
also shown in (a) and (b), respectively. $v_{\rm x}^*$ in (b) is obtained by transferring an extra
eigenvector (with eigenvalue $< 1.04^{-3}$) from the effective null-space of $A_{k n}$ 
to the effective nonsingular space.
}
\end{figure*}

Next, to obtain more rigorously the cutoff eigenvalue for the null-space of $A_{k n}$, we start with 
trial values of the cutoff which are too large (with correspondingly too large trial null-space).
We reduce the cutoff value gradually, changing the division between the 
effective null-space of $A_{k n}$ and the effective nonsingular space of $A_{k n}$, by 
transferring, one by one, eigenvectors from the former to the latter. 
This is straightforward since there is a finite number of discrete eigenvalues. 
At each step we calculate the potentials, $v({\bf r})$ and $w({\bf r})$,  
from Eqs.~(\ref{23m}), (\ref{230}), and (\ref{23}) before and after the shift of the eigenvector respectively. 
We monitor the step change of the potential by calculating the root-square difference,  
\begin{equation}
\Delta v = \sqrt{\int_R d {\bf r} \, \left[ v ({\bf r}) - w ({\bf r}) \right]^2} \, .
\end{equation}
$\int_R$ denotes integration up to a large radius $R$, to prevent divergence of the integral. 
When the eigenvalues are degenerate, we keep or remove the degenerate 
eigenvectors together in each step. 
In Fig. \ref{fig:dv}, we plot the step change of the potential $\Delta v$ versus the eigenvalue 
index $p$ (Ne atom,  basis sets cc-pVTZ and uncontracted cc-pVTZ, $R$=2 a.u.). 
It is shown that the resulting x-OEPs from Eqs.~(\ref{23m}), (\ref{230}), and (\ref{23}) 
change slowly and converge until a truly singular eigenvalue is reached. 
At the step when the latter is transferred from the null-space to the 
nonsingular space, the change of potential $\Delta v$ is abruptly much larger than previous steps. 
We obtain that the smallest nonsingular eigenvalue of $A_{k n}$ is again $g^{\alpha} = 1.04 \times 10^{-3}$.

Comparing the two data sets in Fig.~\ref{fig:dv}, it is evident that the behavior of the step change in the 
potential, versus eigenvalue index $p$ correlates well with the difference of the logarithms of 
consecutive eigenvalues, a simpler procedure that does not require the calculation of the potential in order to 
determine the cutoff. 

The corresponding exchange potential $v_{\rm x}^{\lambda \rightarrow 0}$ is shown in Fig. \ref{f2}(b) 
for cc-pVTZ and uncontracted cc-pVTZ orbital and auxiliary basis sets.
%
We also show $v_{\rm x}^\infty$ (ELP) and $v_{\rm x}^{0.001}$ i.e., the potential for 
$\lambda$ equal to the smallest nonsingular eigenvalue.
Again, $v_{\rm x}^{\lambda \rightarrow 0}$ and $v_{\rm x}^{0.001}$
lie almost on top of each other.
The oscillatory potential $v_{\rm x}^*$ in Fig.~\ref{f2}(b) results from 
transferring an additional eigenvector (with the largest eigenvalue) 
from the null to the nonsingular space of $A_{k n}$. 
The corresponding change of potential is the first large step change $\Delta v$ 
in Fig.~\ref{fig:dv}.

In the inset, as in Fig. \ref{f2}(a), we show the complementary (oscillatory) potentials 
$v^{0}$ and $\bar v$ and their well-behaved sum $v^{\lambda \rightarrow 0}$.
The total energies for $v^{\lambda \rightarrow 0}$ and $v^{0.001}$ are
almost identical and  are $3 \times 10^{-4}$ hartrees higher than the HF energy.
In the last section, the notation ``$\lambda \rightarrow 0$'' 
merely implies use of Eqs.~(\ref{eqv0}),~(\ref{5a}, and (\ref{23m}) to calculate the finite basis 
x-OEP, with the cutoff for the null-space of $A$ determined 
as described. Strictly, $\lambda$ cannot take values lower 
than the small but nonzero eigenvalues in the null-space of $A$. 
 
\section{Conclusion}

We have unveiled nonanalytic behavior of the OEP 
when the density-density response function is truncated with a finite orbital basis set.
We proposed to employ the limiting potential 
$v^{\lambda \rightarrow  0}$, instead of $v^{0}$, 
for the appropriate description of finite-basis OEP.
By using the Uns\"old approximation, 
we derived amended, finite-basis, x-OEP equations~(\ref{23m}), (\ref{230}, and (\ref{23}), 
that determine $v^{\lambda \rightarrow  0}$ completely for any combination of orbital and (large enough) 
auxiliary basis sets.
The Uns\"old approximation amounts to employing an effectively complete orbital 
basis set, i.e. one that includes continuum states.
Our new finite-basis OEP equations do not address the separate problem of how to distinguish 
between the effectively nonsingular and the
singular eigenvalues of the finite basis matrix of the response function $A_{k n}$.
It seems however that with the help of the new finite-basis OEP equations, 
this technical problem may become easier to tackle.


\begin{thebibliography}{99}

\bibitem{sharp}
R.T. Sharp, G.K. Horton, Phys. Rev. {\bf 90}, 317, (1953).

\bibitem{talman}
J.D. Talman, W.F. Shadwick, Phys. Rev. A {\bf 14}, 36 (1976).

\bibitem{kummel2008}
S. K\"ummel, L. Kronik, Rev. Mod. Phys. {\bf 80}, 3, (2008).

\bibitem{bl}
F.A. Bulat, M. Levy, Phys. Rev. A {\bf 80}, 052510, (2009).

\bibitem{hk}
P. Hohenberg, W. Kohn, Phys. Rev. {\bf 136}, B864 (1964).

\bibitem{ks}
W. Kohn, L.J. Sham, Phys. Rev. {\bf 140}, A1133 (1965).

\bibitem{mel}
M. Levy, Proc. Natl. Acad. Sci. U.S.A., {\bf 76}, 6062 (1979).

\bibitem{engel2006}
E. Engel, H. Jiang, Int. J. Quant. Chem. {\bf 106}, 3242, (2006)

\bibitem{rinke2005}
P. Rinke, A. Qteish, J. Neugebauer, C. Freysoldt, M. Scheffler, New Journal of Physics {\bf 7}, 126, (2005).

\bibitem{pnas}
N.I. Gidopoulos, Phys. Rev. A, {\bf 83}, 040502 (2011).

\bibitem{clda}
N.I. Gidopoulos, N. N. Lathiotakis, J. Chem. Phys. (to appear) (2012) 

\bibitem{staroverov2006}
V.N. Staroverov, G.E. Scuseria, E.R. Davidson, J. Chem. Phys. {\bf 124}, 141103, (2006).

\bibitem{hirata2001}
S. Hirata, S. Ivanov, I. Grabowski, R. J. Bartlett, K. Burke, J. D. Talman, 
J. Chem. Phys. {\bf 115}, 1635, (2001).

\bibitem{gorling2008}
A. G\"orling, A. He{\ss}elmann, M. Jones, M. Levy, J. Chem. Phys. {\bf 128}, 104104, (2008).

\bibitem{hess}
A. He{\ss}elmann, A.W. G\"otz, F. Della Sala, A. G\"orling, 
J. Chem. Phys. {\bf 127}, 054102 (2007).

\bibitem{theo_vit}
A.K. Theophilou, V.N. Glushkov, J. Chem. Phys., {\bf 124}, 034105, (2006)

\bibitem{filatov1}
C. Kollmar, M. Filatov, J. Chem. Phys. {\bf 127}, 114104, (2007)

\bibitem{filatov2}
C. Kollmar, M. Filatov, J. Chem. Phys. {\bf 128}, 064101 (2008)

\bibitem{glushkov2009}
V.N. Glushkov, S.I. Fesenko, H.M. Polatoglou, Theor. Chem. Acc. {\bf 124}, 365, (2009)

\bibitem{thb}
T. Heaton-Burgess, F.A. Bulat, W. Yang, Phys. Rev. Lett. {\bf 98}, 256401 (2007).

\bibitem{kli}
J.B. Krieger, Y. Li, G.J. Iafrate, Phys. Rev. A, {\bf 46}, 5453, (1992).

\bibitem{lhf}
F. Della Sala, A. G\"orling, J. Chem. Phys. {\bf 115}, 5718 (2001).

\bibitem{ceda}
M. Gr\"uning, O.V. Gritsenko, E.J. Baerends, J. Chem. Phys. {\bf 116}, 6435 (2002).

\bibitem{elp1}
V.N. Staroverov, G.E. Scuseria, E.R. Davidson, J. Chem. Phys. {\bf 125}, 081104, (2006).

\bibitem{elp2}
A.F. Izmaylov, V.N. Staroverov, G.E. Scuseria, E.R. Davidson, G. Stoltz, E. Canc\`es, 
J. Chem. Phys. {\bf 126}, 0841107, (2007).

\bibitem{ceda0}
A. Uns\"old, Z. Phys. {\bf 43}, 563, (1927). 

\bibitem{savin1}
A. Savin, F. Colonna, M. Allavena, 
J. Chem. Phys. {\bf 115}, 6827 (2001).

\bibitem{nn_tikh}
I. Theophilou, N. N. Lathiotakis, N. I. Gidopoulos, unpublished.


\bibitem{tikhonov} 
W.~H. Press, S. A. Teukolsky, W. T. Vetterling, B. P. Flannery, 
{\it Numerical Recipes: The Art of Scientific Computing}, 3rd ed., 
(Cambridge University Press, New York 2007). 



\end{thebibliography}
\end{document}